\documentclass[aps,prl,twocolumn,showpacs,superscriptaddress,amssymb,amsfonts,amsmath]{revtex4}
\pdfoutput=1 
\usepackage{bm}
\usepackage{graphicx}

\begin{document}

\def \bc {\begin{comment}} \def \ec {end{comment}}
\def\Fbox#1{\vskip1ex\hbox to 8.5cm{\hfil\fboxsep0.3cm\fbox{%
  \parbox{8.0cm}{#1}}\hfil}\vskip1ex\noindent}  


\newcommand{\eq}[1]{(\ref{#1})}
\newcommand{\Eq}[1]{Eq.~(\ref{#1})}
\newcommand{\Eqs}[1]{Eqs.~(\ref{#1})}
\newcommand{\Fig}[1]{Fig.~\ref{#1}}
\newcommand{\Figs}[1]{Figs.~\ref{#1}}
\newcommand{\Sec}[1]{Sec.~\ref{#1}}
\newcommand{\Secs}[1]{Secs.~\ref{#1}}
\newcommand{\Ref}[1]{Ref.~\cite{#1}}
\newcommand{\Refs}[1]{Refs.~\cite{#1}}

\def\be{\begin{equation}}\def\ee{\end{equation}}
\def\bea{\begin{eqnarray}}\def\eea{\end{eqnarray}}
\def\bse{\begin{subequations}}\def\ese{\end{subequations}}
\newcommand{\BE}[1]{\begin{equation}\label{#1}}
\newcommand{\BEA}[1]{\begin{eqnarray}\label{#1}}
\newcommand{\BSE}[1]{\begin{subequations}\label{#1}}

\let \nn  \nonumber  \newcommand{\br}{\\ \nn}
\newcommand{\BR}[1]{\\ \label{#1}}
\def\hf{\frac{1}{2}}
\let \= \equiv \let\*\cdot \let\~\widetilde \let\^\widehat \let\-\overline
\let\p\partial \def\pp {\perp} \def\pl {\parallel}
\def\ort#1{\^{\bf{#1}}}
\def\Trans{^{\scriptscriptstyle{\rm T}}}
\def\x{\ort x} \def\y{\ort y} \def\z{\ort z}
 \def\bn{\bm\nabla} \def\1{\bm1} \def\Tr{{\rm Tr}}
\def\Re{\mbox{  Re}}
\def\<{\left\langle}    \def\>{\right\rangle}
\def\({\left(}          \def\){\right)}
 \def \[ {\left [} \def \] {\right ]}

\renewcommand{\a}{\alpha}\renewcommand{\b}{\beta}\newcommand{\g}{\gamma}
\newcommand{\G} {\Gamma}\renewcommand{\d}{\delta}
\newcommand{\D}{\Delta}\newcommand{\e}{\epsilon}\newcommand{\ve}{\varepsilon}
\newcommand{\E}{\Epsilon}\renewcommand{\o}{\omega} \renewcommand{\O}{\Omega}
\renewcommand{\L}{\Lambda}\renewcommand{\l}{\lambda}
\renewcommand{\t}{\tau}
\def\r{\rho}\def\k{\kappa}
\def\t{\theta } \def\T{\Theta } \def\s{\sigma} \def\S{\Sigma}

\newcommand{\B}[1]{{\bm{#1}}}
\newcommand{\C}[1]{{\mathcal{#1}}}    
\newcommand{\BC}[1]{\bm{\mathcal{#1}}}
\newcommand{\F}[1]{{\mathfrak{#1}}}
\newcommand{\BF}[1]{{\bm{\F {#1}}}}

\renewcommand{\sb}[1]{_{\text {#1}}}  
\renewcommand{\sp}[1]{^{\text {#1}}}  
\newcommand{\Sp}[1]{^{^{\text {#1}}}} 
\def\Sb#1{_{\scriptscriptstyle\rm{#1}}}

\title{Super Stability of Laminar Vortex Flow in Superfluid $^3$He-B }

\author{V.B.~Eltsov}
\affiliation{Low Temperature Laboratory, Aalto University, P.O. Box 15100, FI-00076 AALTO, Finland}
\affiliation{Kapitza Institute for Physical Problems, Kosygina 2, 119334 Moscow,  Russia}

\author{R.~de~Graaf}
\affiliation{Low Temperature Laboratory, Aalto University, P.O. Box 15100, FI-00076 AALTO, Finland}

\author{P.J. Heikkinen}
\affiliation{Low Temperature Laboratory, Aalto University, P.O. Box 15100, FI-00076 AALTO, Finland}

\author{J.J. Hosio}
\affiliation{Low Temperature Laboratory, Aalto University, P.O. Box 15100, FI-00076 AALTO, Finland}

\author{R.~H\"anninen}
\affiliation{Low Temperature Laboratory, Aalto University, P.O. Box 15100, FI-00076 AALTO, Finland}

\author{M.~Krusius}
\affiliation{Low Temperature Laboratory, Aalto University, P.O. Box 15100, FI-00076 AALTO, Finland}

\author{V.S.~L'vov}
\affiliation{Department of Chemical Physics, The Weizmann Institute of
  Science, Rehovot 76100, Israel}

\date{\today}

\begin{abstract}

 Vortex flow remains laminar up to large ($ \sim 10^3$) Reynolds numbers $Re_{\alpha}$ in a cylinder filled with $^3$He-B. This is inferred from NMR measurements and numerical vortex filament calculations where we study the spin up and spin down responses of the superfluid component, after a sudden change in rotation velocity. In normal fluids and in superfluid $^4$He these responses are turbulent. In $^3$He-B the vortex core radius is much larger which reduces both surface pinning and vortex reconnections, the phenomena, which enhance vortex bending and the creation of turbulent tangles. Thus the origin for the greater stability of vortex flow in $^3$He-B is a quantum phenomenon. Only large flow perturbations are found to make the responses turbulent, such as the walls of a cubic container or the presence of invasive measuring probes inside the container.

\end{abstract}
%
\pacs{67.30.hb, 47.15.ki, 02.70.Pt, 67.30.he}

\maketitle 

For centuries the transition to turbulence has been one of the enigmatic unsolved problems of classical hydrodynamics. It is commonly accepted that laminar flow becomes unstable with increasing Reynolds number $Re_{\nu}$ above some case-dependent threshold $Re_\mathrm{th} \sim 1$...$10^3$. This instability results either in secondary vortical flow or (depending on the flow geometry) in intensive turbulent motions with many degrees of freedom. In special geometries, particularly in a circular pipe, laminar flow is assumed asymptotically stable with respect to infinitesimal perturbations at any $Re_{\nu}$ \cite{Hof}, but the critical amplitude of perturbation, which triggers turbulence, decreases rapidly as $\propto 1/Re_{\nu}$ \cite{Mullin}. In practice, normal fluid flows are thus turbulent for  $Re_\nu \gg Re_\mathrm{th}$.

What can be said about the transition to turbulence in the flow of quantized vortices in superfluids? When the temperature $T \rightarrow 0$, dissipation from conventional mechanisms approaches zero and laminar vortex flow is believed to become unstable. This is the current impression about 3-dimensional vortex flow in superfluid $^4$He. Here we show that in superfluid $^3$He-B laminar vortex flow can be stable down to temperatures of $0.2\,T_\mathrm{c}$ in an axially symmetric situation and, to make it turbulent, a strong perturbation is needed.

\textbf{Laminar vortex flow:}--The Reynolds number $Re_{\nu} \equiv VL/\nu$ provides an estimate of the ratio of the nonlinear ($\sim V^2 / L$) and dissipative ($\sim \nu V / L^2$) terms in the Navier-Stokes equation, via the  kinematic viscosity of the fluid, $\nu$, and its  typical velocity and length scales, $V$ and $L$ \cite{Greenspan,Drazin}. Phenomenologically superfluids are described as a mixture of the superfluid and normal components with separate velocities, $\bm{v}_{\rm s}$ and  $\bm{v}_{\rm n}$, densities $\rho\sb s$ and $\rho\sb n$, and viscosities $\nu\sb s\=0$ and $\nu\sb n>0$. The equation for $\bm{v}_{\rm s}(\B r,t)$ is similar to the inviscid Euler equation~\cite{Sonin},
\be \label{SD}     \frac{\p {\B v}\sb s}{\p t}+(\B v \sb s\cdot \B \nabla) {\B v}\sb s+ \B \nabla \mu=  -\alpha'\~{\bm v}\sb s\times \B \omega\sb s+
\alpha~\hat{\B \omega}\times[\B \omega\sb s \times\~{\bm v}\sb s]\,,
\ee
but with additional terms ($\propto \~{\B v}\sb s\=\B v\sb s-\B v\sb n$) on the r.h.s., which describe the dissipative ($\propto \alpha$) and reactive ($\propto \alpha'$) mutual friction between the normal and superfluid components, mediated by the superfluid vorticity $\bm{\omega}_{\rm s} = \bm{\nabla} \times \bm{v}_{\rm s}$. Here $\mu$ is the chemical potential and $\hat{\bm{\omega}} = \bm{\omega}_{\rm s} / \omega_{\rm s}$. Setting $\bm{v}_{\rm n} = 0$,  the ratio of the inertial and dissipative terms in \Eq{SD} can be understood as the \emph{superfluid Reynolds number}~\cite{Finne},
\be
Re_{\alpha} \, (T) \equiv [1- \alpha^{\prime} (T)]\big /  \alpha (T) \approx  1\big /\alpha (T) \,,
\label{2}
\ee
which is independent of velocity, determined only by the temperature dependence of $\a$ and $\a'$, which rapidly decrease towards low temperatures. In the limit $T \rightarrow 0$, $Re_\a\to 1/\a\gg 1$, and the superfluid dynamics is expected to become turbulent.

 \begin{figure*}[t]
 \vspace{-5mm}
 \begin{tabular}{c  c  c} 
$\C A $: signal calibration & $\C B $: vortex flow response & $\C C $: time constants $\tau_{\Downarrow}$ and $\tau_{\Uparrow}$\\
\includegraphics[width=0.35\linewidth,keepaspectratio]{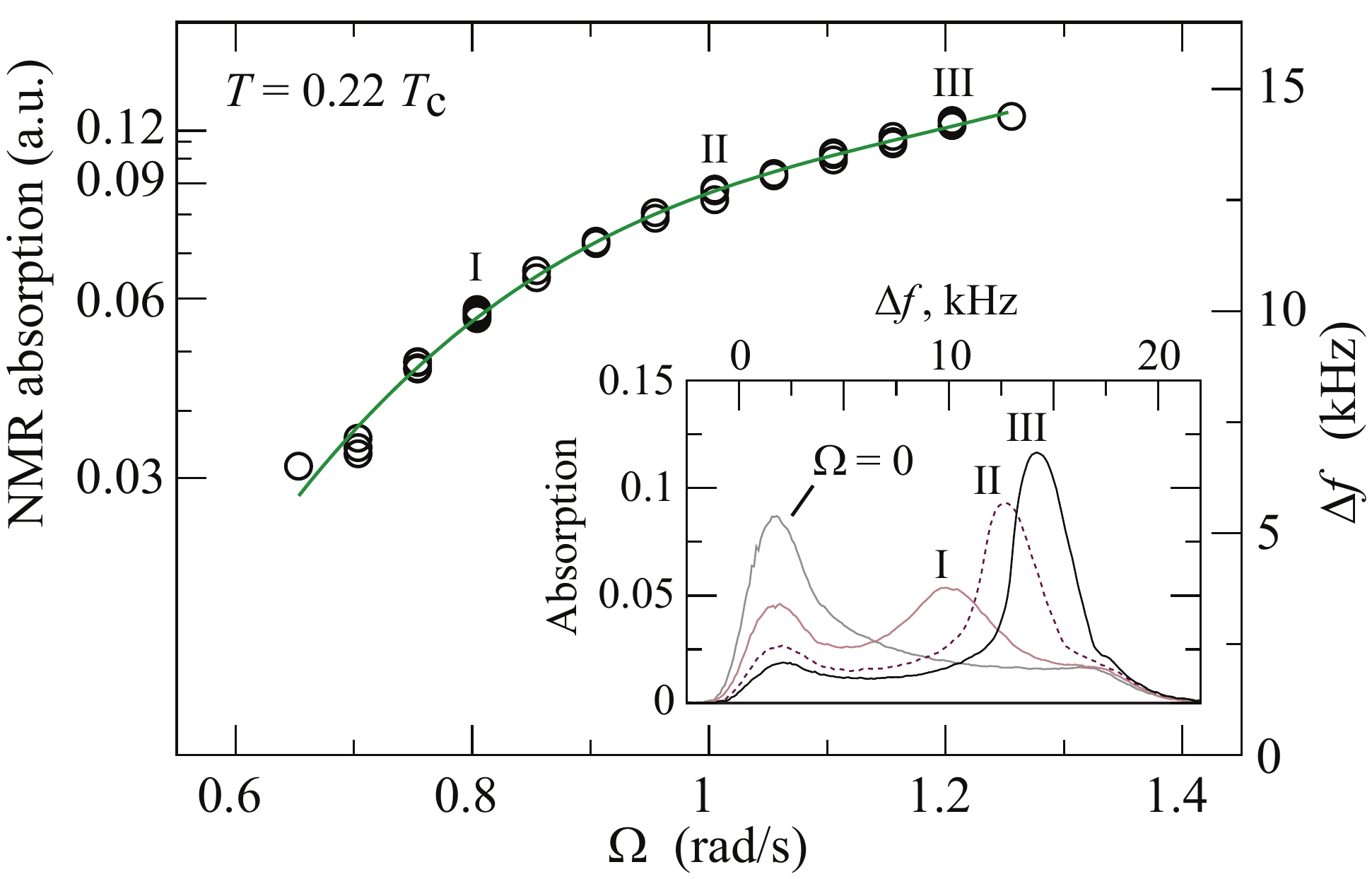} &
\includegraphics[width=0.34\linewidth,keepaspectratio]{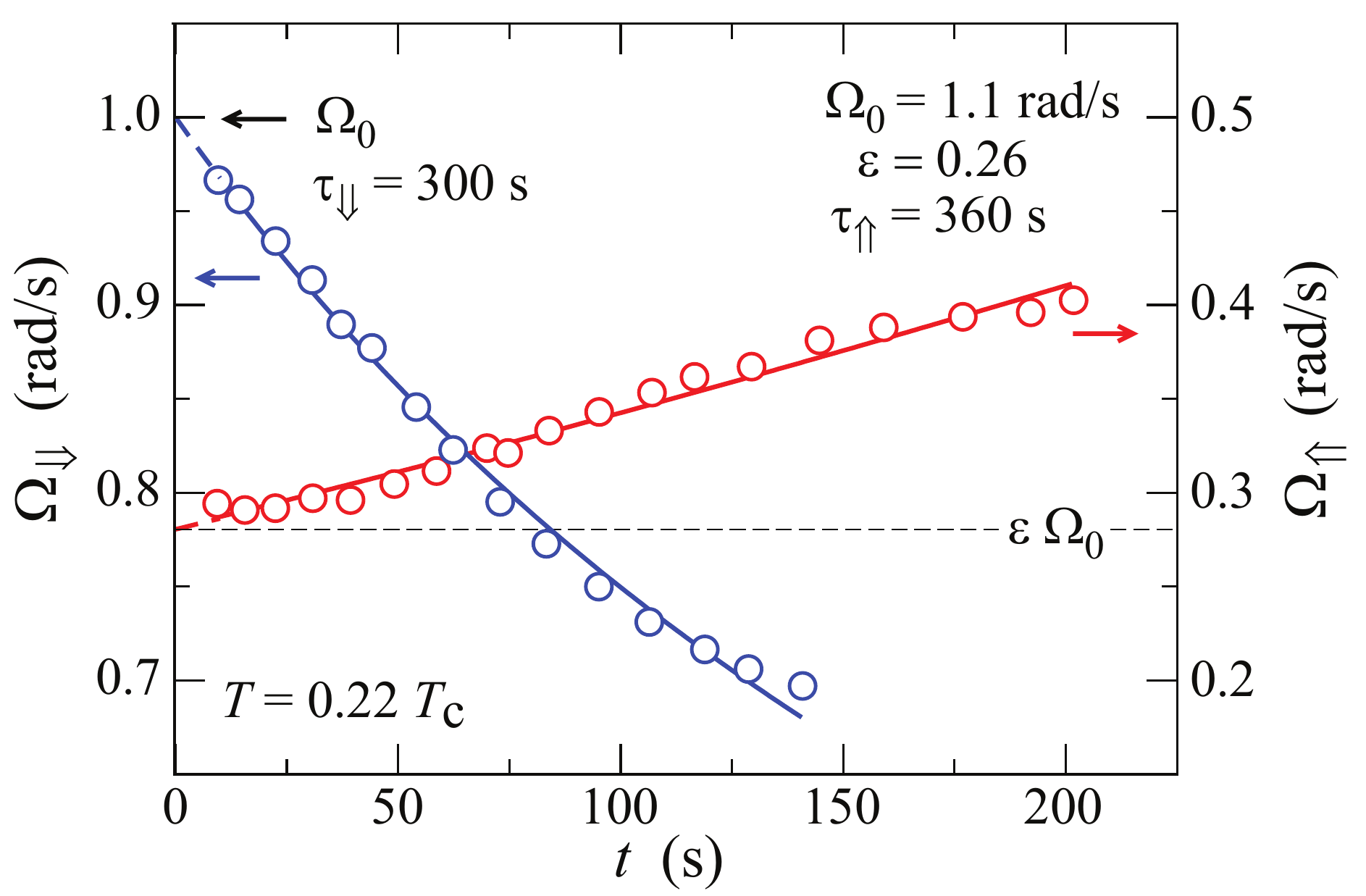} &
\includegraphics[width=0.31\linewidth,keepaspectratio]{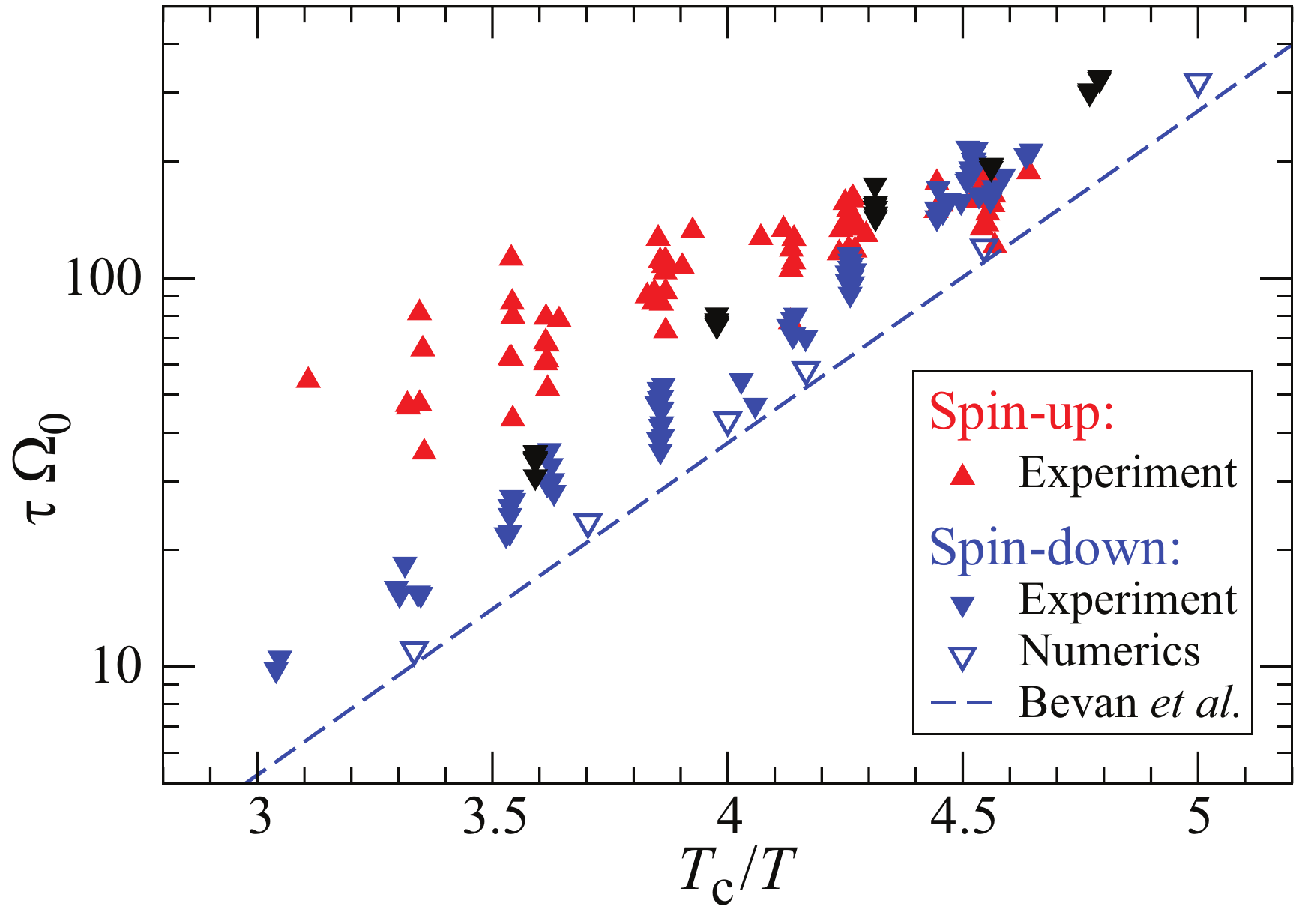} ~~~\\
\end{tabular}
\vspace{-5mm}
\caption{\label{f:1} NMR measurement of spin down and spin up at 29\,bar pressure. \textbf{\textit{(Left)}} Calibration of the cf peak in the NMR absorption spectrum at $0.22\,T_\mathrm{c}$. The height of the peak is plotted on the left vertical axis, its shift $\Delta f = f_\mathrm{cf}- f_\mathrm{L}$ from the Larmor frequency $f_\mathrm{L} = 1.9\,$MHz on the right axis, and the corresponding drive $\O$ at constant rotation in the vortex-free state on the bottom axis. In the insert examples of NMR spectra are shown. \textbf{\textsl{(Middle)}} Spin-down and spin-up responses plotted \textit{versus} time. Using the calibration in the left panel, the heights and/or frequency shifts of the cf peak during its decay have been converted to corresponding $\O$ values. The results have been fitted to Eqs.~(\ref{sol-down}) or (\ref{sol-up}), to extract the characteristic time constants $\tau _{ \Downarrow}$ and $\tau _{ \Uparrow}$ in \Eq{sol-tau}. The extrapolation of the fits to $t=0$ gives $\O_0$ for spin down or $\varepsilon \O_0$ for spin up. \textbf{\textsl{(Right)}} Spin-down and spin-up in terms of $\O_0 \tau = 1/ ( 2 \a)$ \textit{versus} normalized inverse temperature, $T_\mathrm{c} /T$: our measurements (filled triangles) are compared to $1/(2 \a )$ extrapolated from data on $\alpha (T)$ measured above $0.35\,T_\mathrm{c}$ in Ref.~\cite{Bevan} (dashed line), and to the calculated spin down in the cylinder with $\eta = 2^{\circ}$, which uses the mutual friction from Ref.~~\cite{Bevan} (open triangles).
}
\vspace{-5mm}
\end{figure*}

 This is really the case in superfluid $^4$He, where turbulence at the lowest temperatures is ubiquitous in measurements with moving vortices, including so-called  \emph{spin-up} or  \emph{spin-down} measurements, where a sudden change is applied to the angular velocity $\O(t)$ of the rotating cryostat (see~\Ref{4HeSpinUp} for earlier experiments in a cylinder or \Ref{Manchester} in a cube). For simplicity, we restrict the step change to occur from 0 to some $\O$ or from $\O$ to 0. In $^3$He-B the viscous normal component is practically locked to solid-body rotation with the container, while the superfluid component adjusts much slower, coupled only by the small mutual friction from vortex motion with respect to the normal component. Independently of the shape of the container, in experimental work~\cite{4HeSpinUp,Manchester} the turbulent response of the superfluid component has been interpreted as the creation of a turbulent Ekman boundary layer adjacent to the container wall, which expands toward the center of the container. This is similar to what happens in rotating viscous fluids \cite{Greenspan}. In short,  the transition to turbulent vortex flow in superfluid $^4$He generally resembles that in viscous fluids.

In $^3$He-B the situation is different: Our non-invasive NMR measurements show that in a cylinder spin-down is laminar even at $0.2\,T_\mathrm{c}$ and $Re_{\alpha} \sim 10^3$. The same applies for spin up in  a specially prepared situation below $0.3\,T_\mathrm{c}$. In contrast, the response is turbulent when two quartz tuning fork sensors in the middle of the cylinder obstruct cylindrically symmetric flow. In the low temperature limit, where the transport of normal excitations is ballistic, the main difference of $^3$He-B from $^4$He is a much larger vortex core radius, $a\sim 10$...80\,nm. Thus the stability of laminar flow and the transition to turbulence depend on the quantum properties of these two superfluids.

In a long cylinder, oriented parallel to the rotation axis, laminar rotating flow of the superfluid component is solid-body-like, $\langle {\B v}\sb s \rangle = {\B \Omega}(t) \times \B r$, confined to the azimuthal plane. This requires that vortex lines are highly polarized along the cylinder axis and their motion is predominantly 2-dimensional along spirally contracting (spin-up) or expanding (spin-down) trajectories in the azimuthal plane. The motion is obtained from \Eq{SD} which simplifies in terms of its radial dependence to
\be
\frac {d \,\O(t)}{d\, t}= 2 \a \O (t)[\O_0 -\O(t)]\,.\label{3}
\ee
The solutions are for spin down $\O _{ \Downarrow}(t)$ and spin up $\O  _{ \Uparrow} (t)$:
\bse\label{sol}
\BEA{sol-down}
\O _{ \Downarrow}(t)&=& \O_0\big / \big[1+ t/\tau _{ \Downarrow}\big ]\,,
\\ \label{sol-up}
\O _{ \Uparrow}(t)&=& \varepsilon \O_0\big / \big[\varepsilon + (1-\varepsilon) \exp(-t/\tau _{ \Uparrow}\big )]\,,\\ \label{sol-tau}
\tau &=& \tau _{ \Downarrow}(T)= \tau _{ \Uparrow}(T)=[2\a (T)\;  \O_0]^{-1} .
\eea\ese
In spin down the vortex density is $\propto \O_0$ and in spin up $\propto \varepsilon \O_0$ at $t = 0$, when the stable value of drive is reached after the sudden change in rotation.  To conclude, laminar vortex flow displays a monotonic response where $\langle {\B v}\sb s \rangle$ and the total vortex length $L$ smoothly change towards the final state. This is to  be distinguished from a turbulent response where $L$ first builds up to an overshoot and then relaxes faster than in the laminar case.

\textbf{Measurement:}--The vortex flow response is recorded with noninvasive NMR techniques in a smooth-walled quartz cylinder of 6\,mm diameter. The cylinder is aligned along the rotation axis within $\lesssim 1^\circ$ \cite{Front}. The 110\,mm long top NMR section is separated with a flat division wall and a small orifice from the lower part of the cylinder which houses two quartz tuning fork oscillators for temperature measurement \cite{Forks}. Established procedures exist for preparing the upper NMR volume in rotating states with only vortex-free flow \cite{PLTP}. Laminar vortex flow is distinguished by a shift of the NMR absorption to the so-called counterflow (cf) peak in the NMR spectrum \cite{PLTP}. Its height and shift from the Larmor frequency $f_\mathrm{L}$ is a function of the azimuthal large-scale cf and is calibrated experimentally at different values of constant rotation $\O$ in the vortex-free state, as seen in \Fig{f:1}$\C A$.

\begin{figure*}[t]

\begin{tabular}{c  c  c} 
$\C A$: sphere  & $\C B$: cylinder,   inclination angle  $\eta=2^{\circ}$   & $\C C$: cylinders,  $2^{\circ}\leq \eta \leq 70^{\circ} $ \\
\includegraphics[width=0.32\linewidth,keepaspectratio]{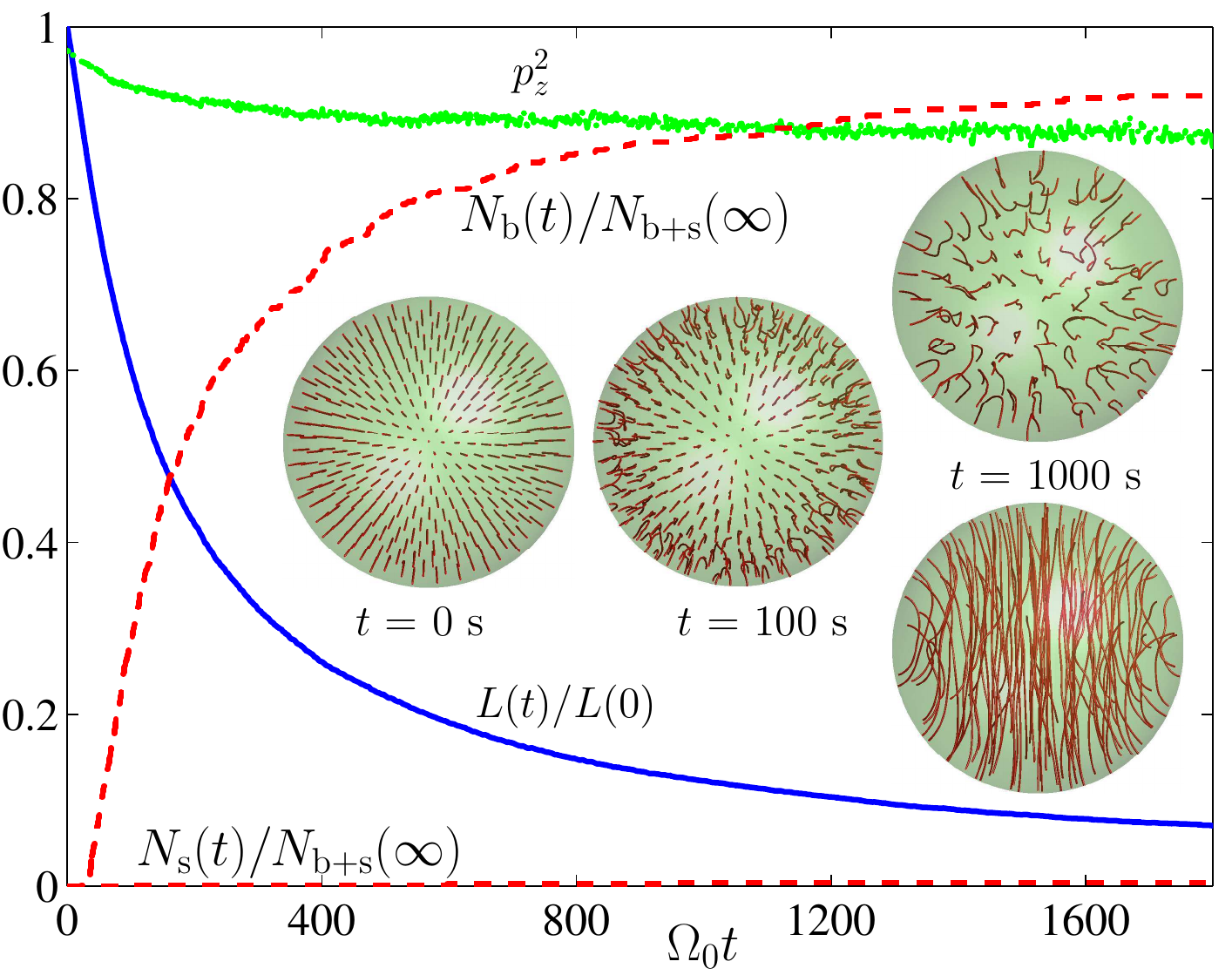} & \includegraphics[width=0.323 \linewidth,keepaspectratio]{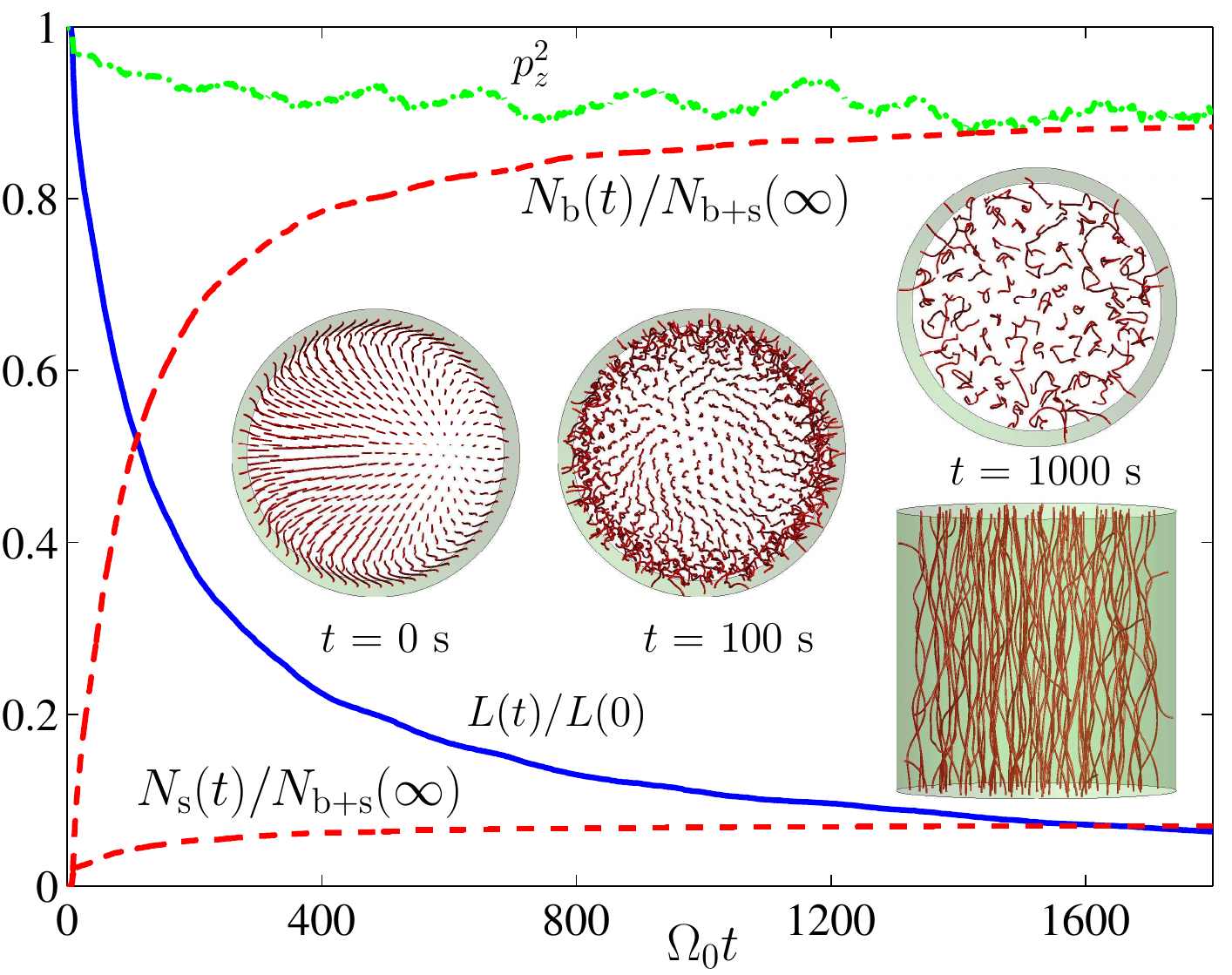} & \includegraphics[width=0.34\linewidth,keepaspectratio]{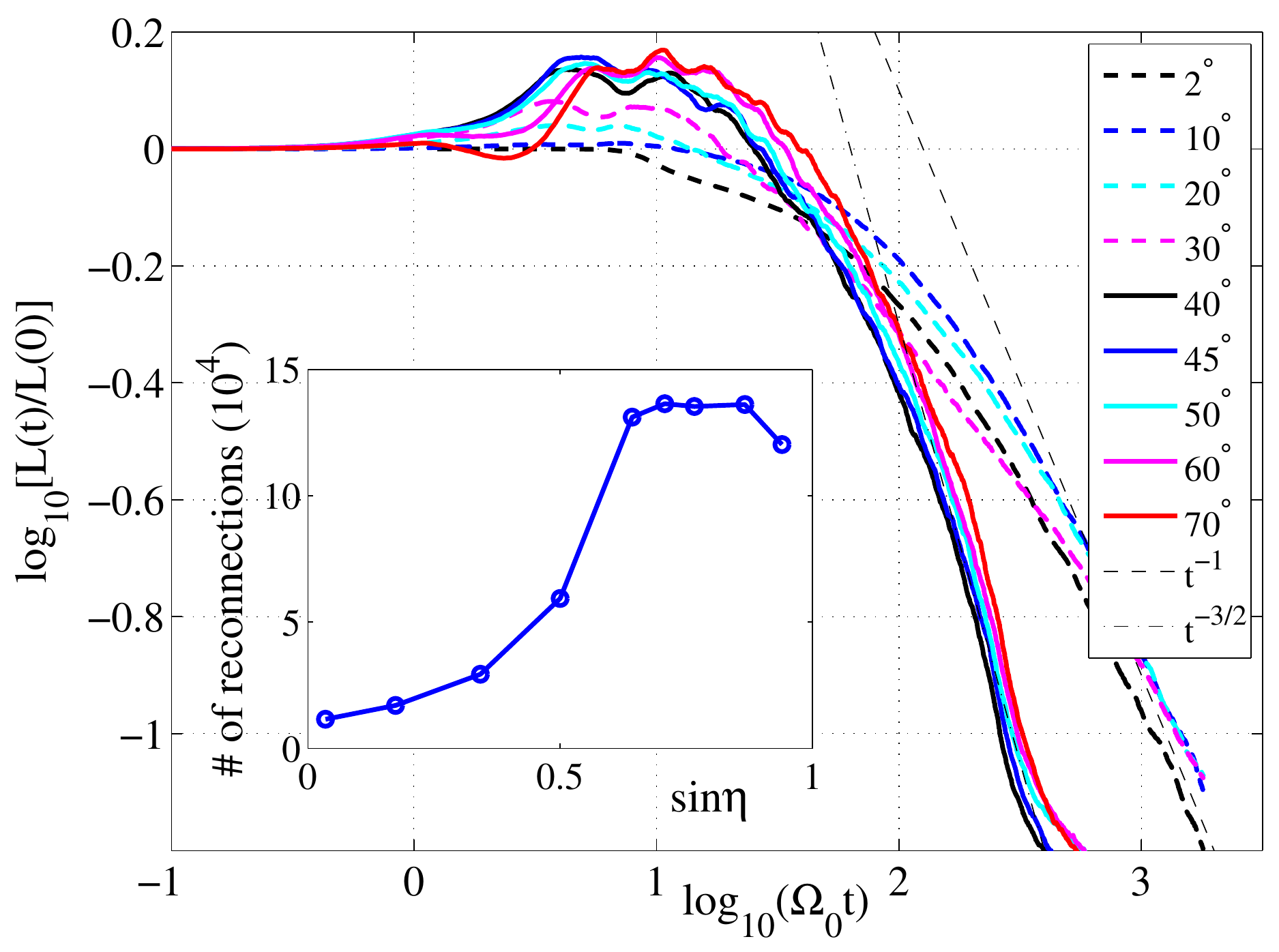} \\ 
\vspace{-5mm}
\end{tabular}
\caption{\label{f:2} Numerical calculations of  spin-down with $\Omega_0=0.5\,$ rad/s, $\alpha = 4.3 \cdot 10^{-3}$, and $\alpha^{\prime} = 9.1 \cdot 10^{-5}$ (which corresponds to $0.22\,T_{\rm c}$ and $29\,$bar). \textbf{\textit{(Left)}} Sphere of radius 3\,mm;  prepared to be initially in the equilibrium vortex state with $\C N(0)= \C N\sb {eq} (\Omega_0) = 325$ vortices; total number of vortex reconnections in the bulk  $N_{\rm b} (\infty) \approx 740$ and with the container surface $N_{\rm s} (\infty) \approx 4$. The normalized curves in the main panel represent: a) total vortex length,  $L(t)/L(0)$ -- solid (blue) line; b) mean vortex polarization along the symmetry axis of the container, $\<p_z^2\>$ -- dotted (green) line; c) cumulative number of the vortex reconnections in the bulk $N\sb b (t)/ [N_{\rm b} (\infty) + N_{\rm s} (\infty)]$ and on the surface $N\sb s (t)/ [N_{\rm b} (\infty) + N_{\rm s} (\infty)]$ -- dashed (red) lines. Inserts: top views  at $t=0$ and  $100\,s$; top and side views  at $t=1000$\,s. \textbf{\textsl{(Middle)}} Cylinder with diameter = height = 6\,mm;  inclined by $\eta = 2^{\circ}$ from the rotation axis;  $\C N(0) = \C N\sb {eq} (\Omega_0) = 413$, $N_{\rm b} (\infty) \approx 10700$, and  $N_{\rm s} (\infty) \approx 860$. \textbf{\textsl{(Right)}} Same cylinder at different inclination angles $\eta$. The time dependence of $L(t)/L(0)$ is plotted on logarithmic scales.  The overall flow behavior is laminar for $\eta \lesssim 30^{\circ}$ and turbulent for $\eta \gtrsim 40^{\circ}$. Insert: total number of all reconnections as a function of the inclination $\eta$.
}\vspace{-5mm}
\end{figure*}

\Fig{f:1}$\C B$ shows an example of the decay of the cf peak during spin down and spin up, as expressed via the equivalent $\O$ from the calibration plot in \Fig{f:1}$\C A$. During the initial sudden change of the drive $\O (t)$ (at $\mid \! \dot{\O} \! \mid = 0.03\,$rad/s$^2$) the cf peak grows continuously and reaches maximum height at $t=0$ when the final constant value of drive is attained. From there on the peak decreases smoothly and monotonicly. If the vortices would lose their polarization because of twisting on the global scale, then flow along the cylinder axis would be created, which would lead to the formation of a new peak close to $f_\mathrm{L}$. However, except for the cf peak from azimuthal flow, no other peaks from large-scale flow have been observed. If a turbulent vortex tangle would fill the cylinder, with a corresponding large loss in vortex polarization and global flow, then all sharp cf peaks would disappear and a broad absorption increase close to $f_\mathrm{L}$ would be detected. Note that the integrated absorption in the NMR spectrum is a constant at a given temperature. Thus a shift of absorption from the cf peak to other frequencies would be seen as \textit{faster} decay of the azimuthal flow. The slow and continuous cf absorption response can only result from smoothly decaying azimuthal global flow and proves the laminar nature of the vortex flow.

\Fig{f:1}$\C C$ shows the characteristic decay time of laminar vortex flow as a function of inverse temperature, as extracted from measurements in the rotation range 0.7...2.5\,rad/s. The data agree well with earlier measurements \cite{Bevan} of  dissipative mutual friction $\a (T)$, considering that temperature measurement in the two experiments also carries  uncertainties. Spin-up times are longer, which is not expected from Eq.~(\ref{sol-tau}), but can be explained by the additional time required for generating new vortices. The spin-up model in Eq.~(\ref{sol-up}) assumes that there exists an instantaneous ample supply of remanent vortices covering homogeneously the cylindrical wall. This requires that a spin-up measurement is performed soon after a previous spin-down run, or else the remanent vortices are reduced to a few and spin-up starts from a localized vortex formation event \cite{PLTP}, followed by axial vortex motion with a propagating front \cite{Front}.

To conclude, the response times in \Fig{f:1}$\C C $ constitute the first observations of laminar spin up and spin down, which remain stable with no additional dissipation beyond the mutual friction from radial vortex motion up to large $Re_{\alpha}$ and possibly even to the limit $Re_{\alpha} \rightarrow \infty$.

\textbf{Numerical simulation:}--To  shed more light on the laminar vortex response, we  use vortex filament calculations \cite{PLTP}. Since vortex creation is problematic in spin-up calculations, only spin down will be considered.

\noindent~\textbullet~\emph{Spherical container}:~\Fig{f:2}$\C A$. The inserts show top and side views of calculated vortex configurations during laminar spin down. These configurations expand in time, with individual vortices moving along spiral trajectories to the equatorial wall, where they finally annihilate as half rings with a radius comparable to the core diameter. During all evolution the configuration is smoothly laminar, with no sign of tangling. This case resembles a decelerating rotating neutron star \cite{Star} and represents a most symmetric flow environment.

\noindent~\textbullet~\emph{Circular cylinder}:~\Fig{f:2}$\C B$.  To mimic a small perturbation from axial symmetry, we  tilted the rotation axis by $\eta = 2^{\circ}$ from the cylinder axis~\cite{Hanninen}. Thus initially at $t=0$ the vortices are slightly inclined with respect to the cylinder. This aberration is small, the vortices are almost rectilinear, of equal length, and their density is constant, $2 \Omega(0)/ \kappa$, as required for solid-body rotation (see top view in the left-most insert; $\kappa = h/(2m_3)$ is the circulation quantum). The insert in the middle shows the top view a little later, when vortices in the outermost ring have partly reconnected at the cylindrical wall, are of reduced length, and are gradually annihilating. Meanwhile vortices inside the outer ring have changed only little in shape.   Much later (top and side views, right-most inserts) the remaining vortices are still evenly distributed and smooth on the scale of the inter-vortex distance $\ell$.

In the main panel of \Fig{f:2}$\C B$ the total vortex length $L(t)$ accurately follows the spin-down dependence of Eq.~(\ref{sol-down}), with $\tau_\Downarrow$ shown in \Fig{f:1}$\C C $ in good agreement with our measurements. The mean polarization remains continuously high, $p_z(t)>0.95$, which applies for all temperatures $\geq 0.20\,T_\mathrm{c}$, but in particular at 0.2...0.3\,$T_\mathrm{c}$  $p_z(t)$ does not change with temperature. Notice that  the total number of reconnections on the cylindrical wall, $N_\mathrm{s} (\infty)$, is roughly twice larger than the initial number of vortices, $\C N(0)$. This is much more than in the case of the sphere and suggests that the main mechanism facilitating vortex annihilation (in the tilted cylinder) is the surface reconnection on the cylindrical wall. Reconnections between vortices in the bulk are 10 times more abundant, but are concentrated to the outermost ring of vortices at early times (where the vortex density is at maximum), as seen in the insert in the middle (at $t = 100\,$s). Thus this numerical example is consistent with our measurements and the claim that laminar spin-down is stable in the bulk volume with respect to finite-size perturbation even at large $Re_\a$, while reconnections and additional dissipation from the annihilation is concentrated in the outermost ring of vortices next to the cylindrical wall.

\noindent~\textbullet~\emph{Tilted cylinder}:~\Fig{f:2}$\C C$.  By varying the inclination $\eta$, we introduce a controllable perturbation. It forces the superfluid to adjust to the elliptically distorted boundary condition and to edge effects at the tilted top and bottom surfaces of the cylinder. This makes it possible to identify a  transition from laminar to turbulent spin down from the time dependence of $L(t)$ at different $\eta$ in the main panel.  When $\eta \lesssim 30^{\circ}$,  $L(t)$ displays no or a minor overshoot, and is followed by a slow fall-off with $t^{-1}$ dependence, typical for laminar flow in Eq.~\eq{sol-down}. When $\eta \gtrsim 40^{\circ}$, the overshoot grows larger and the spin-down decay falls on the fast $t^{-3/2}$ dependence, typical for turbulent flow \cite{nu}.

The overshoot in $L(t)$ arises when vortices reconnect in the bulk volume and form a turbulent tangle. Here the polarization along the rotation axis is largely lost and the kinetic energy of the global azimuthal flow is converted to an increase in vortex density via frequent reconnections. In the insert of \Fig{f:2}$\C C$ the cumulative number of reconnections in the bulk volume is seen to jump up by an order in magnitude, when $\eta \gtrsim 40^{\circ}$, and simultaneously the mean curvature radius of the vortices decreases abruptly. The overshoot is followed by  rapid turbulent decay, displayed by the overall evolution in $L(t)$, although the turbulent tangle may not fill the cylinder homogeneously, but concentrates to its outer regions. It is surprising that such a large inclination as $\eta \sim 30$...40$^\circ$ is required before the overall spin-down response becomes turbulent. This observation underlines the robustness of laminar flow in a cylindrically symmetric environment.

\noindent~\textbullet~\emph{Cubic container:} Similar calculations on a cube show that the spin down response is already turbulent at a low inclination of $\eta = 2^{\circ}$ of the cube faces with respect to the rotation axis \cite{PRL-extension}. The homogeneity and isotropy of the turbulence increase with increasing tilt angle, suggesting that homogeneous turbulence could become possible.

\textbf{Kelvin wave excitations:}--In \Fig{f:2}$\C B$ one can see bending of vortex lines which can be interpreted as Kelvin waves (KWs). To clarify the role of KWs during spin-down in the weakly tilted cylinder, compare the kinetic energy (per unit mass) of laminar rotation $E\Sb {HD}=(\Omega R)^2/4$ with the KW energy $E\Sb {KW}\simeq \Lambda \kappa \Omega (1 - \< p_z^2 \>)\big /2$, where $\Lambda\= \ln (\ell/a)\simeq 10$ in $^3$He-B. One finds
\be \label{rat}
R\Sb {KW / HD}\= E\Sb {KW}\big / E\Sb {HD}\approx 4 \Lambda (1 - \< p_z^2 \>) / \C N \,.
\ee
The polarization is $\< p_z^2 \>\simeq 0.9$ and the number of vortices  $\C N \sim 400$...40 during the evolution shown in \Fig{f:2}$\C B$. Thus the ratio $R\Sb {KW / HD}$ varies between $0.015$ and   $0.15$ and is not changing with temperature, according to the simulation results. This means that KWs practically do not contribute to the total energy of the system. A second argument is derived comparing  dissipation of KWs via mutual friction and via the nonlinear L'vov-Nazarenko KW-energy cascade \cite{LN-2009}. One finds that the cascade may develop only if $\alpha \Lambda \ll (1-\< p_z^2\>)^4$, which is not the case here. We thus conclude that KWs also play no role in the total energy loss, which is consistent with the experimental result.

\textbf{Conclusions:}--In an oriented cylinder vortex flow responses are laminar in the absence of strong flow perturbations and surface pinning. This fact is deduced from the NMR signature of the continuously and monotonicly decaying superfluid counterflow which circulates in the azimuthal plane of the cylinder. The characteristic decay time has the exponential temperature dependence of the $^3$He-B mutual friction dissipation and is exclusively accounted for by radial vortex motion.  The distinctive feature of such laminar flow is a low incidence of vortex reconnections and of Kelvin wave excitations in the bulk volume. As a result, at $0.2\,T_\mathrm{c}$ the dissipation is two orders of magnitude lower than in the  axially inhomogeneous spin up in the form of a propagating turbulent vortex front \cite{Front}. Moreover, in the laminar case dissipation vanishes exponentially in the limit $T \rightarrow 0$, while in the turbulent case it approaches a finite constant value, as confirmed in measurements on the moving front \cite{Front} and on the decay of turbulent vortex tangles \cite{Lancs}.





\begin{references}
\bibitem{nu} Dissipation during the turbulent decay is often expressed as an effective kinematic viscosity $\nu ^\prime/\kappa$ which proves to be $0.02$ (tilted cylinder, $\eta \geq 40^{\circ}$) and $0.03$ (cube, $\eta = 2^{\circ}$).
\end{references}
\end{document}